\documentclass[notitlepage,onecolumn]{article}%
\usepackage{amsmath}
\usepackage{amsfonts}
\usepackage{amssymb}
\usepackage{graphicx}%
\begin{document}


\title{Reply to Comment on ``Deterministic six states protocol for quantum
communication''\ [arXiv:1107.4435v1 [quant-ph] / Phys. Lett. A 374 (2010) 1097]}
\maketitle
\author{
\begin{center}
J.S. Shaari$^{a}$, M. Lucamarini$^{b}$ and Asma' Ahmad Bahari$^{c}$\\$^{a,c}$Faculty of Science, International Islamic University Malaysia (IIUM),
Jalan Sultan Ahmad Shah, Bandar Indera Mahkota,
25200 Kuantan, Pahang,
Malaysia\\
$^{b}$Dipartimento di Fisica, Universit\`a di Camerino, 
Via Madonna delle Carceri, 9
62032, Camerino (MC), Italy
\end{center}
}
\begin{abstract}
We reply to the Comment made in arXiv:1107.4435v1 [quant-ph] (Phys. Lett. A \textbf{374} (2010) 1097) by
noting some erroneous considerations therein resulting in a misleading view of the quantum key distribution protocol in question. We then correct the rates provided for
the Intercept--and-Resend attack and we complete the analysis of Eve's attack
based on a double CNOT gate.
\end{abstract}




In the Comment reported in \cite{1}, the security analysis of the quantum key
distribution protocol (QKD) introduced in \cite{2} and named "6DP" was reviewed and
partially criticized. The main point raised by the Comment is that in the analysis of selected attacks in \cite{2},
6DP's Control Mode uses two qubits from the same basis in contradiction with 6DP's
Encoding Mode where two qubits from different bases were selected for a reliable decoding process. The Comment proceeded with calculations and figures based on the notion that Alice chooses the same bases as that of Bob's preparation for both the qubits in a pair in a control mode.

As we showed in [2] and will further clarify here, it is neither necessary nor efficient to follow such a
prescription in the 6DP and analysis of half the qubit pair is sufficient.  

In Ref. \cite{1} two main attacks from the eavesdropper Eve are analyzed, the
intercept-resend attack (IRA) and the double-CNOT attack (2CNOTA). In the
following, we report the security analysis for these two attacks thus
demonstrating our claims. Moreover we correct a crucial feature of the 2CNOTA
which was not captured in \cite{1}.

\bigskip

We begin with a brief description of 6DP; a two-way QKD protocol based on nonentangled qubits generalizing to the use of 3 mutually unbiased bases (MUB). In 6DP, two qubits are prepared by Bob in two different bases chosen randomly from 3 MUB and sent to and fro to Alice who will encode by executing 1 of 4 unitaries (3 Pauli spin matrices as well as the identity) on each qubit. Bob's sharp measurement of the received qubits would allow him to decode Alice's encoding. Security is achieved by having a control mode where randomly, Alice would with a certain probability make measurements in bases of her choice and through public discussions determine if an eavesdropper was in fact present. 

\bigskip

Arguably the most straightforward eavesdropping strategy against 6DP is the IRA which foresees Eve making measurements of the traveling qubits in the bases of her choice on both the forward and backward path to glean a maximal amount of information of Alice's encoding. 

\bigskip

The IRA was studied in \cite{2} with the main purpose of showing that among
all possible Eve's choices of measuring bases, the one including the same
bases used by Bob, i.e. the canonical $X$, $Y$ and $Z$, is the one which
minimizes the detection probability. The analysis of Control Mode contained in
\cite{2} was limited to only one qubit at a time in the pair of qubits
prepared by Bob. However this does not imply that Eve should necessarily
attack one qubit only, nor does it imply that she should attack both qubits in
the same basis  (this was the allegation made in \cite{1} regarding the analysis in \cite{2}). 

As we described above, the 6DP Encoding Mode is performed by using two
different bases. It is then quite obvious that Eve should attack the protocol
using two different bases, lest her information gain would not be maximized. In
\cite{2} we showed that the minimum probability for Eve to evade
detection\textit{ per single qubit analyzed} by Alice and Bob is $0.5$. Thus the legitimate parties could have inferred the presence of an eavesdropper in either half of the pair as long as their measurement basis coincides; which is certainly more probable than having both qubits measured in coinciding bases. 

The Comment \cite{1} further reported
the probability to evade detection for Eve attacking twice in the same basis as $0.28$, 
claimed for the sake of  comparison to 
what the author had alleged of \cite{2}. We stress that this had never been the case in 
\cite{2} as such an attack has no relevance to the protocol where
Eve would gain less information in doing so. To assume that Eve
performs an attack which cannot provide her with maximal information is quite
arguable and contrary to the principles usually followed in analyzing a quantum
key distribution problem.    

Finally, it is also suggested in Ref. \cite{1} that Eve needs to measure the
traveling qubits in the same basis as Bob's in order to gain from them the
full amount of information encoded by Alice. This is hardly true. In fact, the
measurements performed by Eve during the IRA effectively project the
traveling qubits into her measurement bases, which then are subjected to
Alice's encoding. As noted in \cite{2} (see Fig.1 and following discussion),
Eve's best choice for the bases is any element of the set $\left\{
XY,XZ,YZ,YX,ZX,ZY\right\}  $, composed by all possible different bases
combination, \textit{irrespective of Bob's choice}. In this way, Eve is using
exactly one of the same possible states as that of Bob, thus committing to a faithful decoding of Alice's unitaries and complete information thereof.

\bigskip

Let us now review the analysis of 2CNOTA reported in \cite{1}. For that, it is
useful to write the states prepared by Bob, $\left\vert 0\right\rangle
,\left\vert 1\right\rangle ,\left\vert x_{+}\right\rangle ,\left\vert
x_{-}\right\rangle ,\left\vert y_{+}\right\rangle ,\left\vert y_{-}%
\right\rangle $, as a single state
$a\left\vert 0\right\rangle +b\left\vert 1\right\rangle $
with coefficients $\left(  a,b\right)  $ given by $(1,0)$, $(0,1)$,
$(1/\sqrt{2},1/\sqrt{2})$, $(1/\sqrt{2},-1/\sqrt{2})$, $(1/\sqrt{2},i/\sqrt
{2})$ and $(1/\sqrt{2},-i/\sqrt{2})$ respectively. The global evolution Bob's
state and Eve's ancilla (those with subscript $\ E$) under the 2CNOTA is as
follows:%
\begin{align*}
\mathcal{I}_{\pm}^{z}\left(  a\left\vert 0\right\rangle +b\left\vert
1\right\rangle \right)  \left\vert 0\right\rangle _{E} &  \rightarrow\left(
a\left\vert 0\right\rangle - b\left\vert 1\right\rangle \right)  \left\vert
0\right\rangle _{E},\\
\mathcal{Y}_{\pm}^{z}\left(  a\left\vert 0\right\rangle +b\left\vert
1\right\rangle \right)  \left\vert 0\right\rangle _{E} &  \rightarrow\left(
b\left\vert 1\right\rangle - a\left\vert 0\right\rangle \right)  \left\vert
1\right\rangle _{E},
\end{align*}
where $\mathcal{I}_{+}^{z}\equiv\widehat{C}\left(  I\otimes I\right)
\widehat{C}$, $\mathcal{I}_{-}^{z}\equiv\widehat{C}\left(  Z\otimes I\right)
\widehat{C}$, $\mathcal{Y}_{+}^{z}\equiv\widehat{C}\left(  iY\otimes I\right)
\widehat{C}$ and $\mathcal{Y}_{-}^{z}\equiv\widehat{C}\left(  X\otimes
I\right)  \widehat{C}$ and $\widehat{C}\equiv CNOT$, with Eve's ancilla the
target qubit and Bob's state the control qubit \cite{3}. 
By a \textit{single} execution of this attack Eve only can
distinguish between the sets
\[
S_{11}\equiv\left\{  I,Z\right\}  ,S_{12}\equiv\left\{  iY,X\right\}  ,
\]
but cannot distinguish between the two elements in each set. This means that
by a single execution of the 2CNOTA, Eve cannot access the full 2-bit
information encoded by Alice on each pair of qubits prepared by Bob. This fact
was correctly noted in \cite{1}. What was not noted is that by simply
executing another 2CNOTA (on the other half of the qubit pair), Eve can easily access the full
information. In fact, in the second 2CNOTA, Eve can use an ancilla in the
state $\left\vert x_{+}\right\rangle _{E}$ rather than $\left\vert
0\right\rangle _{E}$ and have her qubit in the position of the control and Bob's as the target \cite{3}.

With this choice, straightforward calculations (not presented here) reveals that the second 2CNOT allows Eve to distinguish between the sets
\[
S_{21}\equiv\left\{  I,X\right\}  ,S_{22}\equiv\left\{  iY,Z\right\}  ,
\]
but not between the two elements in each set, in full analogy with what
happens in the first 2CNOTA (it is just a change of basis). Hence, by
comparing the results obtained in the two executions of the 2CNOTA, Eve can
learn which operation was performed by Alice. For example, if Eve obtains
$S_{12}$ in the first 2CNOTA and $S_{21}$ in the second 2CNOTA, then she will
know that Alice encoded the operation $X$.

Quickly summarizing all of the above, security analysis in 6DP's Control Mode does not compel Eve to measure
twice in the same basis while in the Encoding Mode, she is not
obliged to measure in the same bases as Bob in order to steal full information from
Alice through the IRA. Finally, although a single execution of the 2CNOTA, as
it was originally proposed in \cite{2}, cannot supply Eve with the full amount
of information, a straightforward generalization including two executions of
the 2CNOTA can, thus restituting importance to our original analysis about
this attack. In conclusion, we have shown that the analysis contained in Ref. \cite{1}
contains a number of inaccuracies and in our present Reply we have corrected
them, thus re-establishing the soundness of our original results.
\bigskip
\newline
\noindent{\Large \textbf{Acknowledgement}}
\bigskip
\newline
One of the authors, J.S.S. would like to acknowledge financial
support under the project FRGS0510-122 from the Ministry
of Higher Education's FRGS grant scheme and the University's Research
Management Centre for their assistance and facilities provided.

\end{document}